\documentstyle[graphicx]{elsart}

\def\la{\mathrel{\mathpalette\fun <}}
\def\ga{\mathrel{\mathpalette\fun >}}
\def\fun#1#2{\lower3.6pt\vbox{\baselineskip0pt\lineskip.9pt
  \ialign{$\mathsurround=0pt#1\hfil##\hfil$\crcr#2\crcr\sim\crcr}}}

\begin{document}

\begin{frontmatter}

\title{The Enigma of the Highest Energy Particles of Nature}

\author{G\"unter Sigl}
\address{GReCO, Institut d'Astrophysique de Paris, CNRS,
98bis Boulevard Arago,\\
75014 Paris, France}

\begin{abstract}
Historically cosmic rays have always been at the intersection
of astrophysics with particle physics. This is still and
especially true in current days where experimenters routinely
observe atmospheric showers from particles whose energies reach
macroscopic values up to about 50 Joules. This dwarfs energies
achieved in the laboratory by about eight
orders of magnitude in the detector frame and three orders
of magnitude in the center of mass. While the existence of these
highest energy cosmic rays does not necessarily testify
physics not yet discovered, their macroscopic energies likely
links their origin to the most energetic processes
in the Universe. Explanations range from conventional shock
acceleration to particle physics beyond the Standard Model and
processes taking place at the earliest moments of our Universe.
While motivation for some of the more exotic scenarios may have
diminished by newest data, conventional shock acceleration scenarios
remain to be challenged by the apparent isotropy of cosmic
ray arrival directions which may not be easy to reconcile with
a highly structured and magnetized Universe. Fortunately, many new
experimental activities promise a strong increase of statistics at the highest
energies and a combination with $\gamma-$ray and neutrino astrophysics
will put strong constraints on all these theoretical models.
This short review is far from complete and instead presents a
selection of aspects regarded by the author as interesting and/or promising
for the future.
\end{abstract}

\end{frontmatter}

\section{Introduction}

High energy cosmic ray (CR) particles are shielded
by Earth's atmosphere and reveal their existence on the
ground only by indirect effects such as ionization and
showers of secondary charged particles covering areas up
to many km$^2$ for the highest energy particles. In fact,
in 1912 Victor Hess discovered CRs by measuring ionization from
a balloon~\cite{hess}, and in 1938 Pierre Auger proved the existence of
extensive air showers (EAS) caused by primary particles
with energies above $10^{15}\,$eV by simultaneously observing
the arrival of secondary particles in Geiger counters many meters
apart~\cite{auger_disc}.

After almost 90 years of CR research, their origin
is still an open question, with a degree of uncertainty
increasing with energy~\cite{crbook}: Only below 100 MeV
kinetic energy, where the solar wind shields protons coming
from outside the solar system, the sun must give rise to
the observed proton flux. Above that energy the CR spectrum
exhibits little structure and is approximated
by broken power laws $\propto E^{-\gamma}$:
At the energy $E\simeq4\times 10^{15}\,$eV
called the ``knee'', the flux of particles per area, time, solid angle,
and energy steepens from a power law index $\gamma\simeq2.7$
to one of index $\simeq3.0$. The bulk of the CRs up to at least
that energy is believed to originate within the Milky Way Galaxy. Above the so
called ``ankle'' at $E\simeq5\times10^{18}\,$eV, the spectrum flattens
again to a power law of index $\gamma\simeq2.8$. This latter feature
is often interpreted as a cross over from a steeper Galactic
component, which above the ankle cannot be confined by the Galactic
magnetic field, to a harder component
of extragalactic origin.

Until the 1950s the energies achieved with experiments at accelerators
were lagging behind observed CR energies which explains why many
elementary particles such as the positron, the muon, and the
pion were first discovered in CRs~\cite{battiston}. Today, where the
center of mass (CM) energies observed in collisions with atmospheric nuclei
reach up to a PeV, we have again a similar situation. In addition,
CR interactions in the atmosphere predominantly occur in the extreme
forward direction which allows to probe non-perturbative effects of
the strong interaction. This is complementary to collider experiments
where the detectors can only see interactions with significant
transverse momentum transfer.

Over the last few years, several giant air showers have been
detected both in ground detectors measuring the secondary shower
particles directly in water tanks or scintillation
counters~\cite{haverah,agasa} and in fluorescence telescopes detecting
the nitrogen emission induced by the shower~\cite{fe,hires}.
This confirms the arrival of CRs with energies up to a few hundred
EeV (1 EeV $\equiv 10^{18}\,$eV), corresponding to about 50 Joules.
The existence of such ultra-high energy cosmic rays (UHECRs)
poses a serious challenge for conventional theories of
CR origin based on acceleration of charged particles in powerful
astrophysical objects. The question of the origin of these UHECRs
is, therefore, currently a subject of much theoretical research
as well as experimental efforts. We refer to Ref.~\cite{reviews}
for recent brief reviews, and Ref.~\cite{bs-rev,school} for detailed
reviews.

The problems encountered in trying to explain UHECRs in terms of
``bottom-up'' acceleration mechanisms have been well-documented
in a number of studies; see, e.g., Refs.~\cite{hillas-araa,ssb,norman}.
In summary, apart from energy draining interactions in the source
the maximal UHECR energy is limited by the product
of the accelerator size and the strength of the magnetic field
containing the charged particles to be accelerated, similar to
the situation in man-made accelerators such as at CERN. These
criteria reveal that it is hard to accelerate protons and heavy nuclei
up to the energies observed even in the most powerful astrophysical objects
such as radio galaxies and active galactic nuclei.

\begin{figure}[tbh]
\includegraphics[width=0.95\textwidth]{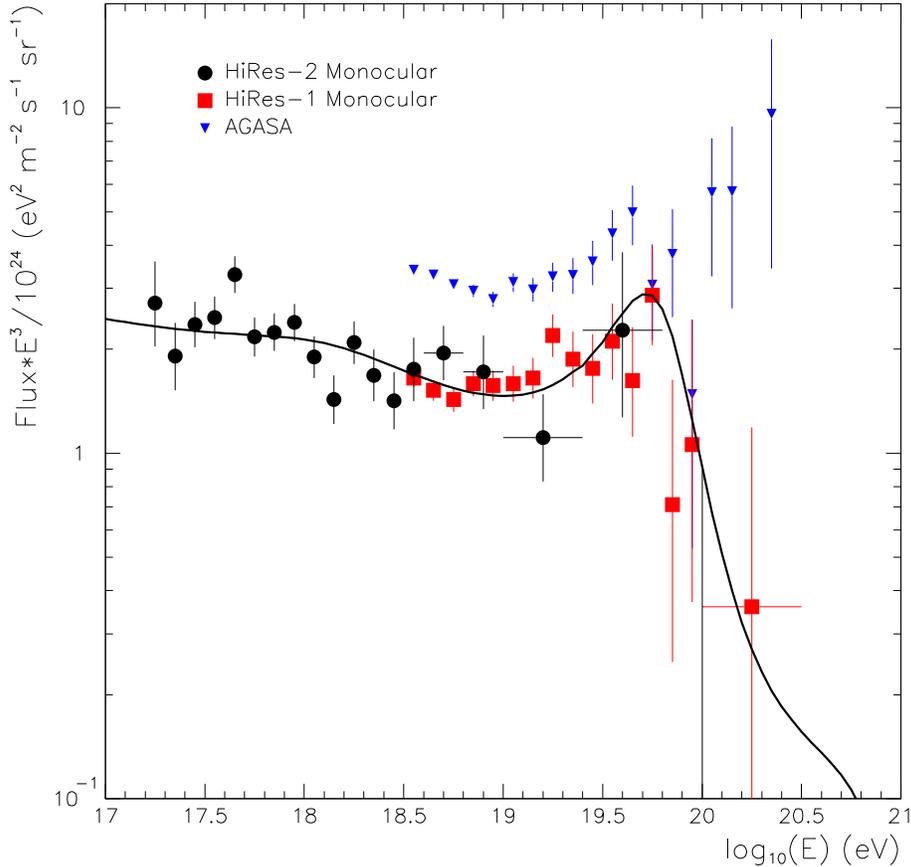}
\caption{UHECR flux as measured by the HiRes-I and HiRes-II
detectors~\cite{hires},
and the AGASA experiment~\cite{agasa}. Also shown is a fit to the data of
a superposition of a galactic and an extragalactic source component.
This figure is from the second reference in Ref.~\cite{hires}.}
\label{fig1}
\end{figure}

In addition, nucleons above
$\simeq70\,$EeV lose energy drastically due to 
photo-pion production on the cosmic microwave background (CMB) 
--- the Greisen-Zatsepin-Kuzmin (GZK) effect~\cite{gzk} --- 
which limits the distance to possible sources to less than
$\simeq100\,$Mpc~\cite{ssb}. Heavy nuclei at these energies
are photo-disintegrated in the CMB within a few Mpc~\cite{heavy}.
If the sources are not too strongly clustered in our cosmologically
local environment, a cut-off in the spectrum above $\simeq70\,$EeV
is therefore expected~\cite{bbo}. However, currently there seems to be
a disagreement between the AGASA ground array~\cite{agasa} which detected
about 8 events above $10^{20}\,$eV, as opposed to about 2 expected
from a cut-off, and the HiRes fluorescence detector~\cite{hires} which
seems consistent with a cut-off~\cite{wb}, see Fig.~\ref{fig1}.
The resolution of this problem
may have to await the completion of the Pierre Auger project~\cite{auger}
which will combine these two complementary detection techniques.

Adding to the problem, there are no  obvious astronomical counterparts
within $\simeq 100\,$Mpc of the Earth~\cite{elb-som,ssb}. At the same
time, no significant large-scale anisotropy has been observed in
UHECR arrival directions above $\simeq10^{18}\,$eV,
whereas there are strong hints for small-scale clustering:
The AGASA experiment has observed five doublets and one triplet within
$2.5^\circ$ out of 57 events above 40 EeV~\cite{agasa}.
This clustering has a chance probability of less than
$1\%$ in the case of an isotropic distribution.

There are currently two possible explanations of these experimental
findings: The first one assumes negligible magnetic deflection. In
this case most of the sources would have to be at cosmological
distances which would explain the absence of nearby counterparts
and the apparent isotropy would indicate that many sources contribute
to the observed flux, where a subset of especially powerful sources
would explain the small-scale clustering~\cite{tt}. This scenario
predicts the confirmation of a GZK cutoff. The second scenario
is most likely more realistic and takes into account the likely
existence of large scale intervening magnetic fields with intensity
$B\sim0.1-1\,\mu$G, correlating with the large scale galaxy distribution.
In this case magnetic deflection would be considerable even at the
highest energies and the observed UHECR flux could be dominated by
relatively few sources within about 100 Mpc. Here, large scale isotropy
could be explained by considerable angular deflection leading to diffusion
up to almost the highest energies and the small scale clustering
could be due to magnetic lensing. This scenario will be discussed in
section~3.

More speculative ways to explain the UHECR observations consist of
circumventing the distance restriction imposed by the GZK effect by
postulating either new particles or new interactions beyond the
Standard Model of particle physics (section 4) or a violation of
the Lorentz symmetry (section 5). These possibilities do not,
however, solve the problem of acceleration to and beyond the observed
energies which has to be solved separately.

In contrast, in the ``top-down'' scenarios, which will be discussed
in section 6, the problem of energetics is
trivially solved. Here, the UHECR particles are the
decay products of some super-massive ``X'' particles of mass
$m_X\gg10^{20}\,$eV, and have energies all the way up to $\sim m_X$. Thus,
no acceleration mechanism is needed. The massive X
particles could be metastable relics of the early Universe
with lifetimes of order of or above the current age of the Universe
or could be released from topological defects (TDs)
that were produced in the early Universe during 
symmetry-breaking phase transitions envisaged in Grand Unified
Theories (GUTs). If the X particles
themselves or their sources cluster similar to dark matter,
the dominant observable  UHECR contribution would come from the
Galactic Halo and absorption would be negligible. This option
seems strongly constrained and may soon be ruled out.

Non-astrophysical solutions of the UHECR problem are of course
in general quite model dependent. In addition, if the existence of the
GZK cutoff will be confirmed, their motivation surely diminishes
in general. On the other hand, even in this case UHECR can be used
to test and constrain new physics beyond the Standard Model,
such as new interactions beyond the reach of terrestrial accelerators
(see section~4), violation of symmetries (see section~5), as well as
Grand Unification and early Universe cosmology, such as the
rate of TD and/or massive particle production in
inflation (see section~6), at energies often inaccessible to
accelerator experiments.

The physics and astrophysics of UHECRs are intimately linked with
the emerging field of neutrino astronomy (for reviews see
Refs.~\cite{nu_review}) as well as with the already
established field of $\gamma-$ray astronomy (for reviews see, e.g.,
Ref.~\cite{gammarev}). Indeed, all
scenarios of UHECR origin, including the top-down models, are severely
constrained by neutrino and $\gamma-$ray observations and limits.
In turn, this linkage has important consequences for theoretical
predictions of fluxes of extragalactic neutrinos above about a TeV
whose detection is a major goal of next-generation
neutrino telescopes: If these neutrinos are
produced as secondaries of protons accelerated in astrophysical
sources and if these protons are not absorbed in the sources,
but rather contribute to the UHECR flux observed, then
the energy content in the neutrino flux can not be higher
than the one in UHECRs, leading to the so called Waxman Bahcall
bound for transparent sources with soft acceleration
spectra~\cite{wb-bound,mpr}.
If one of these assumptions does not apply, such as for acceleration
sources with injection spectra harder than $E^{-2}$ and/or opaque
to nucleons, or in the top-down scenarios where X particle decays
produce much fewer nucleons than $\gamma-$rays and neutrinos,
the Waxman Bahcall bound does not apply, but the neutrino
flux is still constrained by the observed diffuse $\gamma-$ray
flux in the GeV range.

\section{Propagation of Ultra-High Energy Radiation}
Before discussing specific scenarios for the UHECR origin we
give a short account of the interaction and energy loss processes
relevant for the propagation of ultra-high energy cosmic and
$\gamma-$rays, and neutrinos~\cite{sigl}. These processes have
been implemented in various numerical codes used for computing
spectra of all these particles~\cite{code,kkss1,kkss2}. In the
following we assume a flat Universe with a Hubble constant of
$H=70\;{\rm km}\;{\rm sec}^{-1}{\rm Mpc}^{-1}$ and a cosmological
constant $\Omega_\Lambda=0.7$, as favored by current observations.

The relevant nucleon interactions are
pair production by protons ($p\gamma_b\rightarrow p e^- e^+$),
photo-production of single or multiple pions ($N\gamma_b \rightarrow N
\;n\pi$, $n\geq1$), and neutron decay ($n\rightarrow p e^- \bar\nu_e$).
The nucleon threshold energy for single pion production on a
background photon of energy $\varepsilon$ is given by
\begin{equation}
  E_{\rm th}=\frac{m_\pi(m_N+m_\pi/2)}{\varepsilon}\simeq
  6.8\times10^{16}\left(\frac{\varepsilon}{{\rm eV}}\right)^{-1}
  \,{\rm eV}\,.\label{pionprod}
\end{equation}
Typical CMB photon energies are $\varepsilon\sim10^{-3}\,$eV,
leading to the GZK ``cutoff'' at a few tens of EeV
where the nucleon interaction length drops to about $6\,$Mpc.

$\gamma$-rays and electrons/positrons initiate  electromagnetic
(EM) cascades on low energy radiation fields such as the
CMB. The high energy photons undergo electron-positron pair
production (PP; $\gamma \gamma_b \rightarrow e^- e^+$).
At energies below $\sim 10^{14}\,$eV photons interact mainly with 
the universal infrared and optical (IR/O) backgrounds, between
$\sim 10^{14}\,$eV and $\sim100\,$EeV mainly with the CMB,
while above  $\sim100\,$EeV the main target is the universal radio
background (URB). In the Klein-Nishina regime, where the CM energy is
large compared to the electron mass, one of the outgoing particles usually
carries most of the initial energy. This ``leading''
electron (positron) in turn can transfer almost all of its energy to
a background photon via inverse
Compton scattering (ICS; $e \gamma_b \rightarrow e^\prime\gamma$).
EM cascades are driven by this cycle of PP and ICS.
The energy degradation of the ``leading'' particle in this cycle
is slow, whereas the total number of particles grows
exponentially with time. Apart from these interactions of first
order in the EM coupling, triplet pair production (TPP; $e \gamma_b
\rightarrow e e^- e^+$), and double pair production (DPP, $\gamma \gamma_b
\rightarrow e^-e^+e^-e^+$) can play a significant role in the
formation of $\gamma$-ray spectrum in the energy range
$10^8\,{\rm eV} < E < 10^{25}\,$eV. Finally, electrons are
subject to synchrotron losses in large scale extragalactic magnetic fields
(EGMF), whereas for protons synchrotron losses are negligible
outside the sources.

Similarly to photons, ultra-high energy (UHE) neutrinos give rise to
neutrino cascades in the primordial neutrino background via exchange
of W and Z bosons~\cite{zburst1}. Besides the secondary
neutrinos which drive the neutrino cascade, the W and Z decay products
include charged leptons and quarks which in turn feed into the
EM and hadronic channels. Neutrino interactions become
especially significant if the relic neutrinos have masses $m_\nu$
in the eV range and thus constitute hot dark matter, because
the Z boson resonance then occurs at an UHE neutrino energy
$E_{\rm res}=4\times10^{21}({\rm eV}/m_\nu)\,$eV. In fact, the
decay products of this ``Z-burst'' have been
proposed as a significant source of UHECRs~\cite{zburst2}.
The big drawback of this scenario is the need of enormous
primary neutrino fluxes that cannot be produced by known astrophysical
acceleration sources~\cite{kkss1}. Even more exotic top-down type
sources such as X particles decaying mostly into neutrinos
appear ruled out due to a tendency to overproduce
the diffuse GeV $\gamma-$ray flux observed by
EGRET~\cite{egret,kkss2,bko}.

The two major uncertainties in the particle transport are the
intensity and spectrum of the URB for which there exist
no direct measurements in the relevant MHz regime~\cite{Clark,PB},
and the average value of the EGMF. Simulations have been performed
for different assumptions on these uncertainties. A strong URB tends
to suppress the UHE $\gamma$-ray flux by direct absorption
whereas a strong EGMF blocks EM cascading (which otherwise develops
efficiently especially in a low URB) by synchrotron cooling
of the electrons. For the IR/O background we used
the most recent data~\cite{irb} for simulations in the present
paper.

In top-down scenarios, the particle injection spectrum is generally dominated
by the ``primary'' $\gamma$-rays and neutrinos over nucleons. These
primary $\gamma$-rays and neutrinos are produced by the decay of
the primary pions resulting from the hadronization of quarks that come
from the decay of the X particles. In contrast, in acceleration scenarios
the primaries are accelerated protons or nuclei, and
$\gamma$-rays, electrons, and neutrinos are produced as secondaries
from decaying pions that are in turn produced by the interactions
of nucleons with the low energy photon background.

\section{Deflection in Cosmic Magnetic Fields and the Anisotropy Problem}
Magnetic fields are omnipresent in the Universe, but their
origin still lies in the dark~\cite{bt_review}. Best known are
the magnetic fields in galaxies which have strengths of a few
micro Gauss, but there are also some indications for fields correlated
with larger structures such as galaxy clusters~\cite{bo_review}.
Magnetic fields as strong as
$\simeq 1 \mu G$ in sheets and filaments of the large scale galaxy
distribution, such as in our Local Supercluster, are compatible with
existing upper limits on Faraday rotation~\cite{bo_review,ryu,blasi}.
It is also possible that fossil cocoons of former radio galaxies,
so called radio ghosts, contribute to the isotropization of UHECR
arrival directions~\cite{mte}.

Contrary to the case of electrons, for charged hadrons deflection
is more important than synchrotron loss in the EGMF. To get an
impression of typical deflection angles one can characterize the
EGMF by its r.m.s. strength $B$ and a coherence length $l_c$.
If we neglect energy loss processes for the moment, then
the r.m.s. deflection angle over a distance $r$ in such a field
is $\theta(E,r)\simeq(2rl_c/9)^{1/2}/r_L$~\cite{wm}, where the Larmor
radius of a particle of charge $Ze$ and energy $E$ is
$r_L\simeq E/(ZeB)$. In numbers this reads
\begin{equation}
  \theta(E,r)\simeq0.8^\circ\,
  Z\left(\frac{E}{10^{20}\,{\rm eV}}\right)^{-1}
  \left(\frac{r}{10\,{\rm Mpc}}\right)^{1/2}
  \left(\frac{l_c}{1\,{\rm Mpc}}\right)^{1/2}
  \left(\frac{B}{10^{-9}\,{\rm G}}\right)\,,\label{deflec}
\end{equation}
for $r\ga l_c$. This expression makes it immediately obvious
why a magnetized Local Supercluster with fields of fractions
of micro Gauss prevents a direct assignment of sources in
the arrival directions of observed UHECRs; the deflection
expected is many tens of degrees even at the highest energies.
This goes along with a time delay $\tau(E,r)\simeq r\theta(E,d)^2/4
\simeq1.5\times10^3\,Z^2(E/10^{20}\,{\rm eV})^{-2}\\
(r/10\,{\rm Mpc})^{2}(l_c/{\rm Mpc})(B/10^{-9}\,{\rm G})^2\,$yr
which can be millions of years. A source visible in UHECRs today
could therefore be optically invisible since many models involving,
for example, active galaxies as UHECR accelerators, predict
variability on shorter time scales.

On the other hand, an EGMF of this size could explain the
observed large scale UHECR isotropy by diffusion and the small-scale
clustering by magnetic lensing, even if most of the sources are
relatively nearby. In fact, numerical simulations
of nucleons in a simplified scenario where the UHECR source density
in the Local Supercluster is idealized as a Gaussian sheet of a few
Mpc thickness and about 20 Mpc length containing a magnetic
field with a Kolmogorov spectrum and an r.m.s. strength $B$
proportional to the same profile have lead to the following
result: About 10 sources in the Local Supercluster and a
maximal field strength of $\simeq0.3\mu\,$G lead to arrival
direction multi-pole moments and autocorrelation functions
consistent with the AGASA data~\cite{is}.

\begin{figure}[ht]
\includegraphics[width=\textwidth,clip=true]{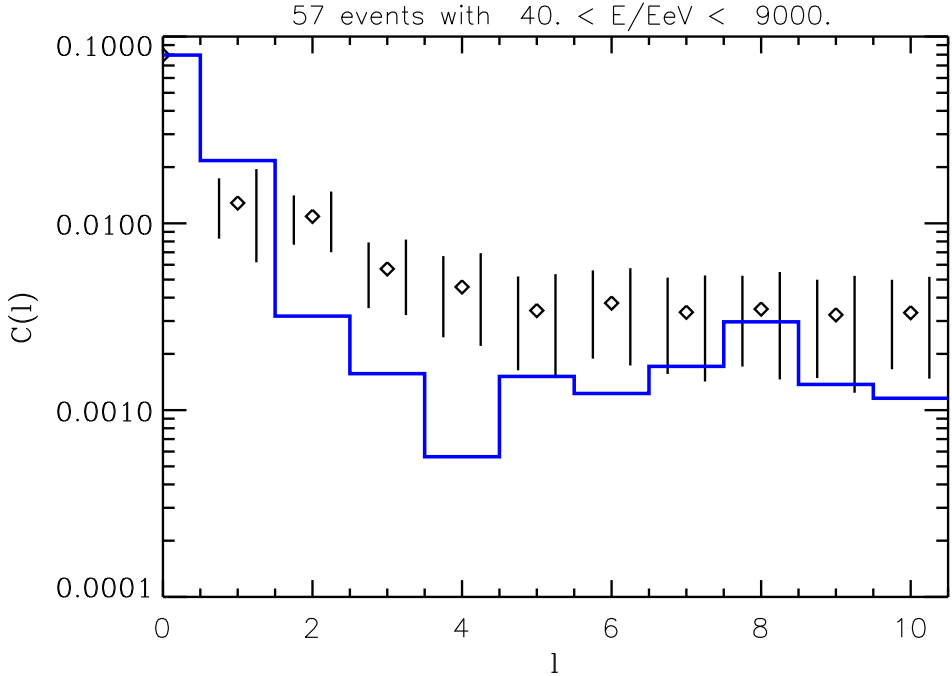}
\caption[...]{The angular power spectrum $C(l)$
as a function of multi-pole $l$, obtained for the AGASA
exposure function, for $N=57$ events observed above 40 EeV, sampled
from 12 simulated configurations of 100 sources in the simulation
box of Ref.~\cite{miniati}, emitting a proton spectrum $\propto E^{-2.4}$
up to $10^{21}\,$eV. The magnetic field strength at the observer
is $\simeq0.13\mu\,$G. The diamonds indicate
the average over 12 realizations, and the left and right error
bars represent the statistical and total (including cosmic variance
due to different realizations)
error, respectively, see text for explanations. The histogram
represents the AGASA data. The overall likelihood significance is
$\simeq0.15$.}
\label{fig2}
\end{figure}

\begin{figure}[ht]
\includegraphics[width=\textwidth,clip=true]{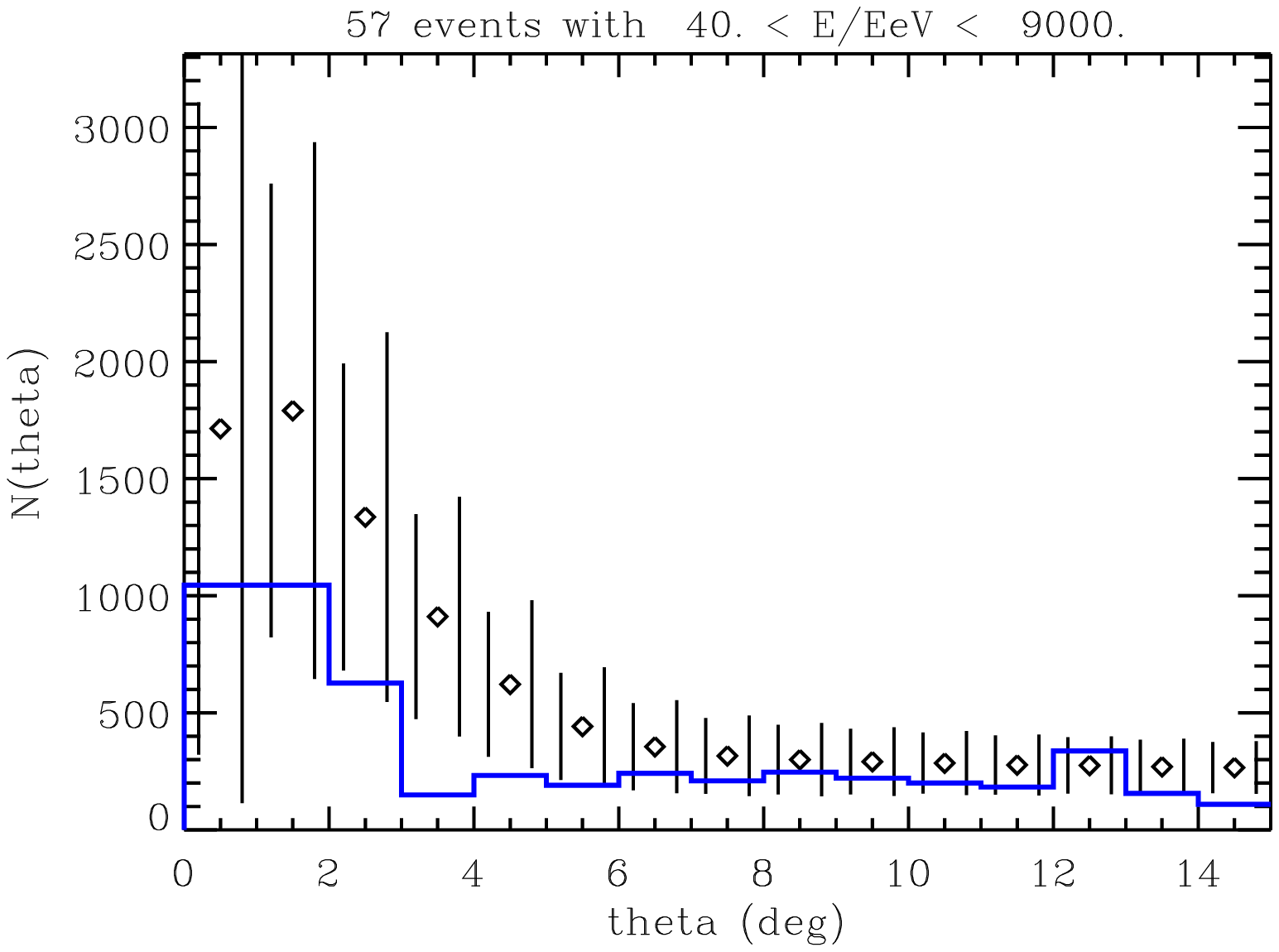}
\caption[...]{As Fig.~\ref{fig2}, but for the angular correlation
function $N(\theta)$ as a function of angular distance $\theta$,
using a bin size of $\Delta\theta=1^\circ$. The overall likelihood
significance is $\simeq0.63$.}
\label{fig3}
\end{figure}

Most recent results indicate that the situation is similar in more
realistic models of the Local Supercluster. We have used the
$(50\,{\rm Mpc}^3$ box of a large scale structure
simulation~\cite{miniati} where the
magnetic field was followed passively and normalized to
the known strengths of a few micro Gauss in the center of
galaxy clusters. In this simulated box we selected as observer
a place typical for our neighborhood in the Local Supercluster,
and chose randomly a certain number of sources with probability
proportional to the local baryon density. For each such
configuration 5000 nucleon trajectories originating from the
sources (assumed to emit equal spectra and power) and reaching
the observer were computed numerically,
taking into account energy losses and deflection in the
magnetic field in the box. For each realization these 5000 trajectories
were used to construct arrival direction probability distributions
taking into account the solid-angle dependent exposure function
for the respective experiment and folding over the angular resolution.
From these distributions
mock data sets consisting of $N$ observed events were
dialed. For each such mock data set or for the real data
set we then obtained estimators for the spherical harmonic coefficients
$C(l)$ and the autocorrelation function $N(\theta)$. The
estimator for $C(l)$ is defined as
\begin{equation}
  C(l)=\frac{1}{2l+1}\frac{1}{{\cal N}^2}
  \sum_{m=-l}^l\left(\sum_{i=1}^N\frac{1}{\omega_i}Y_{lm}(u^i)
  \right)^2\,,\label{cl}
\end{equation}
where $\omega_i$ is the total experimental exposure
at arrival direction $u^i$, ${\cal N}=\sum_{i=1}^{N}1/\omega_i$
is the sum of the weights $1/\omega_i$, and
$Y_{lm}(u^i)$ is the real-valued spherical harmonics function
taken at direction $u^i$. The estimator for $N(\theta)$ is defined as
\begin{equation}
N(\theta)=\frac{1}{2S(\theta)}\sum_{j \neq i}
\left\{\begin{array}{ll}
1 & \mbox {if $\theta_{ij}$ is in same bin as $\theta$}\\
0 & \mbox{otherwise}
\end{array}\right\}\,,
\label{auto}
\end{equation}
and $S(\theta)$ is the solid angle size of the corresponding bin.

The different mock data sets in the various realizations
yield the statistical distributions of $C(l)$ and $N(\theta)$.
One defines the average over all mock data sets and realizations
as well as two errors: The smaller error (shown to the left of
the average in Figs.~\ref{fig2} and~\ref{fig3}) is the statistical
error, i.e. the fluctuations due to the finite number $N$ of observed
events, averaged over all realizations, while the larger error
(shown to the right of the average in Figs.~\ref{fig2} and~\ref{fig3})
is the ``total error'', i.e. the statistical error plus the
cosmic variance, in other words, the fluctuations due to finite
number of events and the variation between different realizations
of observer and source positions. For a given data set a $\chi^2$
summed over all bins can be defined relative to the mean and variance
of $C(l)$ and $N(\theta)$ simulated for a given model. From
this one can obtain an overall likelihood for the
consistency of a given real data set with a given model
by counting the fraction of simulated mock data sets with a $\chi^2$
larger than for the real data.

\begin{figure}[ht]
\includegraphics[width=\textwidth,clip=true]{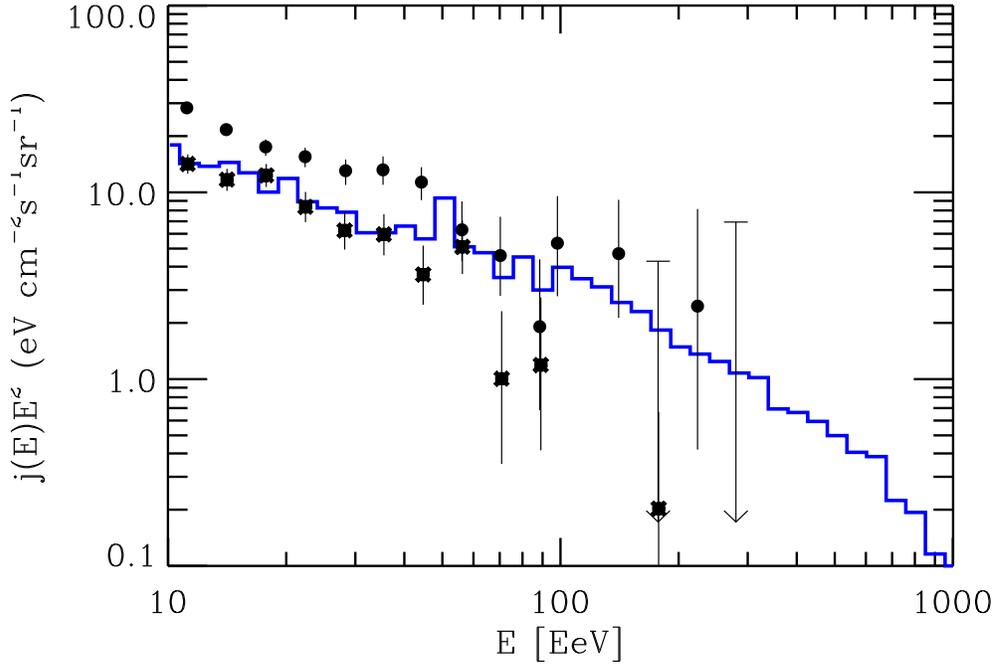}
\caption[...]{Predicted spectrum observable by AGASA for the scenario
discussed in the text and shown
in Figs.~\ref{fig2} and~\ref{fig3}, averaged over 12 realizations,
as compared to the AGASA (dots) and HiRes-I (stars) data.}
\label{fig4}
\end{figure}

We find that as long as the observer is surrounded by magnetic
fields above about $0.1\mu$G, which is possible but not obvious for our
environment, roughly 10 or more sources lead to multi-poles
and autocorrelations marginally consistent with present data which are
limited to the Northern hemisphere, although consistency
of large scale multi-poles is somewhat worse than for more extended
EGMFs. An example for 100
sources is shown in Figs.~\ref{fig2} and~\ref{fig3}~\cite{ems}.
A proton injection spectrum roughly $\propto E^{-2.4}$ extending
up to $\simeq10^{21}\,$eV can reproduce
the sub-GZK spectrum and tends to predict a spectrum somewhere
between the AGASA and HiRes observations above GZK energies,
see Fig.~\ref{fig4}. In contrast, if the observer is in a region
of EGMF field strength much smaller than $\simeq0.1\mu$G, the UHECR
sky distribution reflects the highly structured large scale
galaxy distribution and is thus inconsistent with the observed
isotropy.

We also find that sources outside our Local Supercluster
do not contribute significantly to the observable flux if the observer
is immersed in magnetic fields above about $0.1\mu$G and if
sources reside in magnetized clusters and super-clusters:
For particles above the GZK cutoff this is because sources outside
the Local Supercluster are beyond the GZK distance. Sub-GZK particles
are mainly contained in the magnetized environment of their sources
and thus exhibit a much higher local over-density than
their sources. Further, the suppressed flux of low energy particles
leaving their environment is largely kept out of our Local
Supercluster by its own magnetic field~\cite{is}. Both effects can be
understood qualitatively by matching the flux $j(E)$ in the unmagnetized
region with the diffusive flux $-D(E){\bf\nabla}n(E,{\bf r})$ in terms
of the diffusion constant $D(E)$ and the density $n(E,{\bf r})$ of
particles of energy $E$ which shows that the density gradient always
points to the source. A significant contribution from sources at
cosmological distance would
require magnetic fields in the nano Gauss range, and even then
would be unlikely to explain the observed clustering.

\begin{figure}[ht]
\includegraphics[width=\textwidth,clip=true]{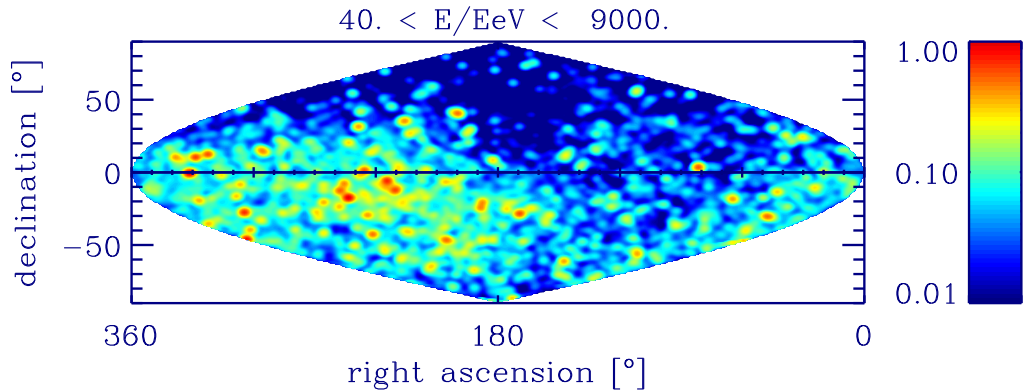}
\caption[...]{The UHECR arrival direction distribution above 40 EeV
in terrestrial coordinates for the scenario of Figs.~\ref{fig2}--\ref{fig4},
averaged over all 12 realizations
of 5000 trajectories each. The color scale represents the integral
flux per solid angle. The pixel size is $1^{\circ}$ and the image
has been convolved to an angular resolution of $2.4^{\circ}$
corresponding to the approximate AGASA angular resolution.}
\label{fig5}
\end{figure}

The confidence levels that can be obtained with this method for
specific models of our local magnetic and UHECR source neighborhood
will greatly increase with the increase of data anticipated from future
experiments. Full sky coverage alone will play an important role
in this context as many scenarios predict large dipoles for the
UHECR distribution. This is the case for the specific simulation
box on which Figs.~\ref{fig2}--\ref{fig4} are based, as demonstrated
in Fig.~\ref{fig5}. Whereas current northern hemisphere data are consistent
with this scenario at the $\simeq1.5\,$sigma level, a comparable
or larger exposure in the southern hemisphere would be
sufficient in this case to find a dipole at several sigma confidence
level.

Modeling our cosmic neighborhood and simulating UHECR
propagation in this environment will therefore become more and more
important in the coming years. This will also have to include the
effects of the Galactic magnetic field and an extension to a possible
heavy component of nuclei. For first steps in this direction see,
e.g. Ref.~\cite{ames} and Ref.~\cite{bils}, respectively.

\section{New Primary Particles and Interactions}

A possible way around the problem of missing counterparts
within acceleration scenarios is to propose primary
particles whose range is not limited by interactions with
the CMB. Within the Standard Model the only candidate is the neutrino,
whereas in extensions of the Standard Model one could think of
new neutrals such as axions or stable supersymmetric elementary
particles. Such options are mostly ruled out by the tension
between the necessity of a small EM coupling to avoid the GZK cutoff and
a large hadronic coupling to ensure normal air showers~\cite{ggs}.
Also suggested have been
new neutral hadronic bound states of light gluinos with
quarks and gluons, so-called R-hadrons that are heavier than
nucleons, and therefore have a higher GZK threshold~\cite{cfk},
as can be seen from Eq.~(\ref{pionprod}).
Since this too seems to be disfavored by accelerator
constraints~\cite{gluino} we will here focus on neutrinos.

In both the neutrino and new neutral stable particle scenario
the particle propagating over extragalactic distances would have
to be produced as a secondary in interactions of a primary proton
that is accelerated in a powerful active galactic nucleus
which can, in contrast to the case of EAS induced by nucleons,
nuclei, or $\gamma-$rays,
be located at high redshift. Consequently, these scenarios predict
a correlation between primary arrival directions
and high redshift sources. In fact, possible evidence
for a correlation of UHECR arrival directions with compact radio
quasars and BL-Lac objects, some of them possibly too far away to
be consistent with the GZK effect, was recently reported~\cite{bllac}.
The main challenge in these correlation studies is the choice
of physically meaningful source selection criteria and the avoidance
of a posteriori statistical effects. However, a moderate increase
in the observed number of events will most likely confirm or
rule out the correlation hypothesis. Note, however, that
these scenarios require the primary proton to be accelerated
up to at least $10^{21}\,$eV, demanding a very powerful
astrophysical accelerator.

\subsection{New Neutrino Interactions}

Neutrino primaries have the advantage of being well established
particles. However, within the Standard Model their interaction cross
section with nucleons, whose charged current part can
be parametrized by~\cite{gqrs}
\begin{equation}
\sigma^{SM}_{\nu N}(E)\simeq2.36\times10^{-32}(E/10^{19}
  \,{\rm eV})^{0.363}\,{\rm cm}^2\,,\label{cross}
\end{equation}
for $10^{16}\,{\rm eV}\la E\la10^{21}\,{\rm eV}$, falls short by about
five orders of magnitude to produce air showers as they are observed.
However, it has been suggested that the neutrino-nucleon
cross section, $\sigma_{\nu N}$, can be enhanced by new
physics beyond the electroweak scale in the
CM frame, or above about a PeV in the nucleon rest frame.
Neutrino induced air showers may therefore rather directly
probe new physics beyond the electroweak scale.

One possibility consists of a large increase
in the number of degrees of freedom above the electroweak 
scale~\cite{kovesi-domokos}. A specific instance
of this idea appears in theories with $n$ additional large
compact dimensions and a quantum gravity scale $M_{4+n}\sim\,$TeV
that has recently received much attention in the literature~\cite{tev-qg}
because it provides an alternative solution to the hierarchy problem
in grand unifications of gauge interactions without a need
of supersymmetry. One of the largest contributions to the neutrino-nucleon
cross section turns out to be the production on the brane representing
our world of microscopic black holes which extend into
the extra dimensions. The production of
compact branes, completely wrapped around the extra dimensions,
may provide even larger contributions~\cite{aco}. The cross sections
can be larger than in the Standard model one by up to a factor
$\sim100$ for reasonable parameters~\cite{fs}. However, this is not
sufficient to explain the observed UHECR events~\cite{kp}.

\begin{figure}[ht]
\includegraphics[width=0.7\textwidth,clip=true,angle=270]{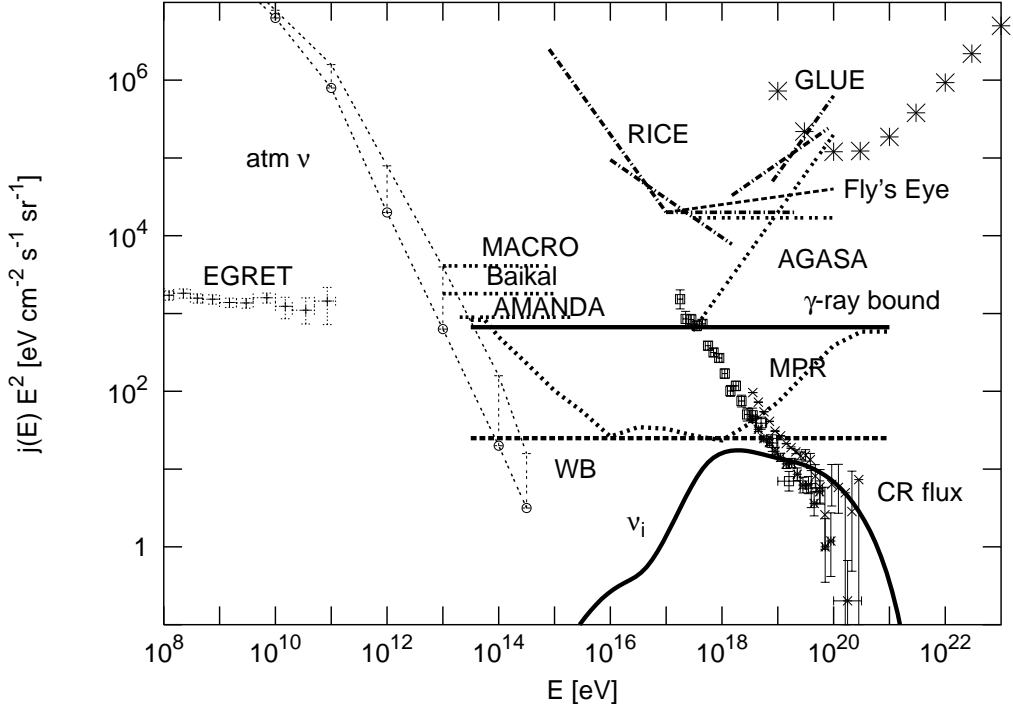}
\caption[...]{\cite{dima}~Cosmogenic neutrino
flux per flavor (thick line, assuming maximal mixing among all flavors)
produced by primary proton flux from the AGASA~\cite{agasa} and
HiRes~\cite{hires} experiments above $3\times10^{18}\,$eV (error bars).
The UHECR sources were assumed to inject a proton spectrum
$\propto E^{-2}$ up to $10^{22}\,$eV with luminosity $\propto(1+z)^3$ up to
$z=2$. For comparison shown are the
atmospheric neutrino flux~\cite{lipari}, as well as existing upper
limits on the diffuse neutrino fluxes from MACRO~\cite{macro},
AMANDA~\cite{amanda_limit},
BAIKAL~\cite{baikal_limit}, AGASA~\cite{agasa_nu}, the
Fly's Eye~\cite{baltrusaitis} and RICE~\cite{rice} experiments, and
the limit obtained with the Goldstone radio telescope (GLUE)~\cite{glue},
as indicated. Theoretical neutrino flux upper limits based on the
$\gamma-$ray bound derived from the EGRET diffuse $\gamma-$ray
flux (shown on left part between 30 MeV and 100 GeV)~\cite{egret},
the MPR limit for optically thin  sources~\cite{mpr}, and the WB limit for
AGN-like redshift evolution~\cite{wb-bound} are also shown.}
\label{fig6}
\end{figure}

However, the UHECR data can be used to put constraints on cross sections
satisfying $\sigma_{\nu N}(E\ga10^{19}\,{\rm eV})
\la10^{-27}\,{\rm cm}^2$. Particles with such cross
sections would give rise to horizontal air showers which have
not yet been observed. Resulting upper limits on their fluxes
assuming the Standard Model cross section Eq.~(\ref{cross}) are
shown in Fig.~\ref{fig6}. Comparison with the ``cosmogenic'' neutrino flux
produced by UHECRs interacting with the CMB then results in upper
limits on the cross section which are about a factor 1000 larger
than Eq.~(\ref{cross}) in the energy range between $\simeq10^{17}\,$eV
and $\simeq10^{19}\,$eV~\cite{mr,tol,afgs}. The projected sensitivity of
future experiments shown in Fig.~\ref{fig7} below indicate that these limits
could be lowered down to the Standard Model one~\cite{afgs}. In case
of a detection of penetrating events the degeneracy of the cross
section with the unknown neutrino flux could be broken by comparing the rates
of horizontal air showers with the ones of Earth skimming events~\cite{kw}.

The CM energy reached by an UHECR nucleon of energy $E$ interacting
with an atmospheric nucleon at rest is
$\sqrt s\simeq0.4(E/10^{20}\,{\rm eV})\,$PeV.
Since the starting points in the atmosphere of observed EAS at these energies
are consistent with normal hadronic cross sections, scenarios
predicting dramatic increases or saturation of cross sections
in the TeV range can probably be excluded. It is generally expected
that new physics appears at TeV scales, and dramatic effects are
possible when this scale is related to a true fundamental
gravity scale. There are, for example, string-inspired scenarios with
infinite-volume extra dimensions and a string scale $M_{\rm s}\simeq
M_{\rm eff}^2/M_{\rm Pl}\sim(M_{\rm eff}/TeV)^210^{-3}\,$eV, where
$M_{\rm eff}$ is the energy scale where cross sections would
undergo a Hagedorn-like saturation~\cite{dgln,dgs}, and
$M_{\rm Pl}\simeq1.72\times10^{18}\,$GeV is the reduced Planck mass.
The UHECR argument would then require $M_{\rm eff}\ga\,$PeV
and thus $M_{\rm s}\ga100\,$eV.
However, in case of at least two infinite extra dimensions
gravity would appear higher-dimensional,
i.e. decreasing with a higher power of distance than Newtonian gravity
on scales $r\ga M_{\rm Pl}/M_{\rm s}^2\simeq1\,{\rm pc}
\,(M_{\rm s}/100{\rm eV})^{-2}$. The above bound would thus imply
modifications of gravity on parsec scales which seems
phenomenologically inviable. This is an example where UHECR
observations can constrain models not constrained yet by
collider physics due to the lower CM energies reached. Of course,
UHECR observations are much more indirect and thus are only
sensitive to dramatic effects such as Hagedorn saturation
and macroscopic cross sections~\cite{sigl1}.

\section{Violation of Lorentz Invariance}
The most elegant solution to the problem of apparently missing nearby
sources of UHECRs and for their putative correlation with high redshift
sources would be to speculate that the GZK effect does not exist theoretically.
A number of authors pointed out~\cite{vli_others,cg} that this may be possible
by allowing violation of Lorentz invariance
(VLI) by a tiny amount that is consistent with all current
experiments. At a purely theoretical level, several quantum gravity models
including some based on string theories do in fact predict non-trivial
modifications of space-time symmetries that also imply VLI at extremely
short distances (or equivalently at extremely high energies); see e.g., 
Ref.~\cite{amelino-piran} and references therein. These theories 
are, however, not yet in forms definite enough to allow precise
quantitative predictions of the exact form of the possible VLI. 
Current formulations of the effects of a possible VLI on high energy
particle interactions relevant in the context of UHECR, therefore, adopt a 
phenomenological approach in which the form of the possible
VLI is parametrized in various
ways. VLI generally implies the existence of a universal preferred frame
which is usually identified with the frame that is
comoving with the expansion of the Universe, in which the CMB
is isotropic. 

A direct way of introducing VLI is through a modification
of the standard {\it dispersion relation},
$E^2-p^2=m^2$, between energy $E$ and momentum $p=|\vec{p}|$ of particles, $m$
being the invariant mass of the particle. Currently there is no 
unique way of parameterizing the possible modification of this relation in
a Lorentz non-invariant theory. We discuss here a parametrization
of the modified dispersion relation which covers most of the qualitative
cases discussed in the literature and, for certain parameter values,
allows to completely evade the GZK limit,
\begin{equation}
  E^2-p^2-m^2\simeq-2dE^2-\xi\frac{E^3}{M_{\rm Pl}}-
  \zeta\frac{E^4}{M^2_{\rm Pl}}\,.\label{vli1}
\end{equation}
Here, the Planck mass $M_{\rm Pl}$ characterizes non-renormalizable effects
with dimensionless coefficients $\xi$ and $\zeta$, and the dimensionless
constant $d$ exemplifies VLI effects due to renormalizable terms
in the Lagrangian. The standard Lorentz invariant dispersion relation
is recovered in the limit $\xi,\zeta,d\to0$. In critical string theory,
effects second order in $M^{-1}_{\rm Pl}$, $\zeta\neq0$, can be induced
due to quantum gravity effects.
The constants $d\neq0$ can break Lorentz invariance spontaneously when
certain Lorentz tensors $c_{\mu\nu}$ have couplings to fermions of the form
$d_{\mu\nu}\bar\psi\gamma^\mu\partial^\nu\psi$, and acquire
vacuum expectation values of the form
$\langle d_{\mu\nu}\rangle=d\delta_\mu^0\delta_\nu^0$~\cite{ck}. Effects of
first order in $M^{-1}_{\rm Pl}$, $\xi\neq0$, are possible, for example, in
non-critical Liouville string theory due to recoiling D-branes~\cite{emn}.

Now, consider the GZK photo-pion production process in
which a nucleon of energy $E$, momentum $p$ and mass $m_N$ collides
head-on with a CMB photon of energy $\epsilon$ producing a pion and a
recoiling nucleon. The threshold initial momentum of the nucleon for this
process according to standard Lorentz invariant kinematics is 
\begin{equation}
p_{{\rm th},0}=(m_{\pi}^{2}+2m_{\pi}m_N)/4\epsilon\,,\label{gzk-th} 
\end{equation}
where $m_\pi$ and $m_N$ are the pion and nucleon masses, respectively.
Assuming exact energy-momentum conservation but using the modified
dispersion relation given above, in the
ultra-relativistic regime $m\ll p\ll M$, and neglecting sub-leading terms,
the new nucleon threshold momentum $p_{\rm th}$ under the modified
dispersion relation Eq.~(\ref{vli1}) for $d=0$ satisfies~\cite{aloisio} 
\begin{equation} 
-\beta x^4-\alpha x^3 + x - 1 = 0\,,
\label{vli2} 
\end{equation}
where $x=p_{\rm th}/p_{{\rm th},0}$, and 
\begin{eqnarray} 
\alpha&=&\frac{2\xi p_{{\rm th},0}^{3}}{(m_{\pi}^{2}+2m_\pi m_N)M_{\rm Pl}}
 \frac{m_\pi m_N}{(m_\pi + m_N)^2}\label{vli3}\\
\beta&=&\frac{3\zeta p_{{\rm th},0}^{4}}{2(m_{\pi}^{2}+2m_\pi m_N)M^2_{\rm Pl}}
 \frac{m_\pi m_N}{(m_\pi + m_N)^2}\,.\nonumber
\end{eqnarray} 
One can show that the same modified dispersion
relation Eq.~(\ref{vli1}) leads to the same condition Eq.~(\ref{vli2})
for absorption of high energy gamma rays through $e^+e^-$ pair production
on the infrared, microwave or radio backgrounds, if one substitutes
$p_{{\rm th},0}=m^2_e/\epsilon$,
$\alpha=\xi p_{{\rm th},0}^3/(8m^2_e M_{\rm Pl})$,
$\beta=3\zeta p_{{\rm th},0}^4/(16m^2_e M^2_{\rm Pl})$, where $m_e$
is the electron mass.

If $\xi,\zeta\simeq1$, there is no real positive solution of
Eq.~(\ref{vli2}), implying that the GZK process does not take place and
consequently the GZK cutoff effect disappears completely. Thus UHE
nucleons and/or photons will be able to reach Earth from any distance. 
On the other hand, if future UHECR data confirm the presence of a GZK
cutoff at some energy then that would imply upper limits on the
couplings $\xi$ and $\zeta$, thus probing specific
Lorentz non-invariant theories. If $p_{\rm th}\simeq p_{{\rm th},0}$,
one could conclude from Eq.~(\ref{vli2}) that $\alpha,\beta\la1$,
which translates into $|\xi|\la10^{-13}$ for the first order effects,
and $|\zeta|\la10^{-6}$ for the second order
effects, $\xi=0$~\cite{aloisio}. Confirmation of a cut-off for
TeV photons with next-generation $\gamma-$ray observatories
would lead to somewhat weaker constraints~\cite{tev}.

In addition, the non-renormalizable terms in the dispersion relation
Eq.~(\ref{vli1}) imply a change in the group velocity which for the
first-order term leads to time delays over distances $r$ given by
\begin{equation}
  \Delta t\simeq\xi r\frac{E}{M_{\rm Pl}}\simeq\xi
  \left(\frac{r}{100\,{\rm Mpc}}\right)\left(\frac{E}{{\rm TeV}}\right)
  \,{\rm sec}\,.\label{delay} 
\end{equation}
For $|\xi|\sim1$ such time delays could be measurable, for example,
by fitting the arrival times of $\gamma-$rays arriving from
$\gamma-$ray bursts to the predicted energy dependence.

We mention that if VLI is due to modification of the
space-time structure expected in some theories of quantum gravity,
for example, then the strict energy-momentum conservation assumed in the above
discussion, which requires space-time translation invariance, 
is not guaranteed in general, and then the calculation of the modified
particle interaction thresholds becomes highly non-trivial and
non-obvious. Also, it is possible that a Lorentz non-invariant
theory while giving a  modified dispersion relation also imposes
additional kinematic structures such as a modified law of addition
of momenta. Indeed, Ref.~\cite{amelino-piran} gives an example of
a so-called $\kappa$-Minkowski non-commutative space-time in which the modified
dispersion relation has the same form as in Eq.~(\ref{vli1}) but there is also
a modified momentum addition rule which compensates for the effect of the
modified dispersion relation on the particle interaction thresholds
discussed above leaving the threshold momentum unaffected and consequently  
the GZK problem unsolved. In scenarios where the relativity of
inertial frames is preserved by a non-linear representation of the
Poincar\'e group, thresholds are in general significantly modified
only if the effective mass scale $M_{\rm Pl}/xi$ is of the order
of the unmodified threshold energy in the laboratory frame~\cite{ms}.

VLI by dimensionless terms such as $d$ in Eq.~(\ref{vli1}) has been
considered in Ref.~\cite{cg}. These terms are obtained by adding
renormalizable terms that break Lorentz invariance to the Standard
Model Lagrangian. The dimensionless terms can be interpreted as
a change of the maximal particle velocity
$v_{\rm max}=\partial E/\partial p|_{E,p\gg m}\simeq1-d$. At a fixed
energy $E$ one has the correspondence $d\to(\xi/2)(E/M_{\rm Pl})+
(\zeta/2)(E/M_{\rm Pl})^2$, as can be seen from Eq.~(\ref{vli1}).
The above values for $\xi$ and $\zeta$
influencing the GZK effect then translate into values
$|d|\ga10^{-24}$.

There are
several other fascinating effects of allowing a small VLI, some of which
are relevant for the question of origin and propagation of UHECR, and
the resulting constraints on VLI parameters from cosmic ray observations  
are often more stringent than the corresponding laboratory limits;
for more details, see Ref.~\cite{cg} and~\cite{jlm}.

\section{Top-Down Scenarios}
The difficulties of bottom-up acceleration scenarios discussed
earlier motivated the proposal of the ``top-down'' scenarios,
where UHECRs, instead of being accelerated, are the decay products
of certain ``X'' particles of mass close to the GUT scale. Such
particles can be produced in basically two ways: If they are very
short lived, as usually expected in many GUTs, they have to be
produced continuously. The only way this can be achieved is
by emission from TDs left over from cosmological
phase transitions that may have occurred in the early Universe at
temperatures close to the GUT scale, possibly during reheating
after inflation. TDs
necessarily occur between regions that are causally disconnected, such
that the orientation of the order parameter
associated with the phase transition, can not be communicated
between these regions and consequently will adopt different
values. Examples are cosmic strings, magnetic monopoles, and
domain walls. The defect density is thus given by the particle horizon
in the early Universe.
The defects are topologically stable, but time dependent motion
leads to the emission of particles of mass comparable to the
temperature at which the phase transition took place. The
associated phase transition can also occur during reheating
after inflation.

Alternatively, instead of being released from TDs, X particles
may have been produced directly in the early Universe and,
due to some unknown symmetries, have a very
long lifetime comparable to the age of the Universe.
In contrast to Weakly-Interacting Massive Particles (WIMPS)
below a few hundred TeV which are the usual dark matter
candidates motivated by, for example, supersymmetry and can
be produced by thermal freeze out, such super-heavy X particles
have to be produced non-thermally (see Ref.~\cite{kuz-tak}
for a review). In all these cases, such particles,
also called ``WIMPZILLAs'', would contribute to the dark matter
and their decays could still contribute to UHECR fluxes today,
with an anisotropy pattern that reflects the dark matter
distribution in the halo of our Galaxy. However, these scenarios
predict the absence of the GZK cutoff~\cite{kb} and thus will
be ruled out if the existence of the GZK cutoff is confirmed~\cite{hires}.

It is interesting to note that one of the prime motivations
of the inflationary paradigm was to dilute excessive production
of ``dangerous relics'' such as TDs and
super-heavy stable particles. However, such objects can be
produced right after inflation during reheating
in cosmologically interesting abundances, and with a mass scale
roughly given by the inflationary scale which in turn
is fixed by the CMB anisotropies to
$\sim10^{13}\,$GeV~\cite{kuz-tak}. The reader will realize that
this mass scale is somewhat above the highest energies
observed in CRs, which implies that the decay products of
these primordial relics could well have something to do with
UHECRs which therefore can probe such scenarios!

For dimensional reasons the spatially averaged X particle
injection rate can only
depend on the mass scale $m_X$ and on cosmic time $t$ in the
combination
\begin{equation}
  \dot n_X(t)=\kappa m_X^p t^{-4+p}\,,\label{dotnx}
\end{equation}
where $\kappa$ and $p$ are dimensionless constants whose
value depend on the specific top-down scenario~\cite{BHS}.
For example, the case $p=1$ is representative of scenarios
involving release of X particles from TDs,
such as ordinary cosmic strings~\cite{BR}, monopoles connected
by strings, so-called ``necklaces''~\cite{BV} and magnetic 
monopoles~\cite{BS}. This can be easily seen as follows:
The energy density $\rho_s$ in a network of defects has to scale
roughly as the critical density, $\rho_s\propto\rho_{\rm crit}\propto
t^{-2}$, where $t$ is cosmic time, otherwise the defects
would either start to over-close the Universe, or end up
having a negligible contribution to the total energy
density. In order to maintain this scaling, the defect
network has to release energy with a rate given by
$\dot\rho_s=-a\rho_s/t\propto t^{-3}$, where $a=1$ in
the radiation dominated era, and $a=2/3$ during matter
domination. If most of this energy goes into emission
of X particles, then typically $\kappa\sim{\cal O}(1)$.
In the numerical simulations presented below, it was
assumed that the X particles are non-relativistic at decay.

The X particles could be gauge bosons, Higgs bosons, super-heavy fermions,
etc.~depending on the specific GUT. They would have
a mass $m_X$ comparable to the symmetry breaking scale and would
decay into leptons and/or quarks of roughly
comparable energy. The quarks interact strongly and 
hadronize into nucleons ($N$s) and pions, the latter
decaying in turn into $\gamma$-rays, electrons, and neutrinos. 
Given the X particle production rate, $dn_X/dt$, the effective
injection spectrum of particle species $a$ ($a=\gamma,N,e^\pm,\nu$) 
via the hadronic channel can be
written as $(dn_X/dt)(2/m_X)(dN_a/dx)$,
where $x \equiv 2E/m_X$, and $dN_a/dx$ is the relevant
fragmentation function (FF).

The FFs are governed by quantum chromo dynamics (QCD). At energies
reachable at accelerators the FFs have been measured in some detail.
At the GUT scale energies relevant for top-down scenarios,
however, extrapolations based on QCD have to be employed to obtain
the FFs. In our calculations we adopt the Local Parton Hadron Duality
(LPHD) approximation~\cite{detal} according to which the total
hadronic FF, $dN_h/dx$, is taken to be proportional to the spectrum
of the partons (quarks/gluons) in the parton cascade (which is initiated
by the quark through perturbative QCD processes) after evolving the parton
cascade to a stage where the typical transverse momentum transfer in the
QCD cascading processes has come down to $\sim R^{-1}\sim$ few hundred 
MeV, where $R$ is a typical hadron size. The parton spectrum is obtained
from solutions of the standard QCD evolution equations in modified leading
logarithmic approximation (MLLA) which provides good fits to accelerator
data at LEP energies~\cite{detal}. Within the LPHD hypothesis, the pions
and nucleons after hadronization have essentially the same spectrum. 
The LPHD does not, however, fix the relative abundance of pions and
nucleons after hadronization. Motivated by accelerator data, we assume
the nucleon content $f_N$ of the hadrons to be $\simeq10$\%, and the
rest pions distributed equally among the three charge states.
Recent work on FFs goes beyond MLLA by solving the QCD evolution equations
numerically and also includes the supersymmetric degrees of freedom at
energies above the supersymmetry breaking scale~\cite{frag}. The
main difference in the results seems to be a nucleon-to-pion ratio
that can be significantly higher at large $x$ values for the
extremely high energies of interest here~\cite{frag}. However, the
situation is not completely settled yet and the flux predictions
are not very sensitive to such differences in the injection spectrum.

\subsection{Predicted Fluxes}

\begin{figure}[ht]
\includegraphics[width=0.75\textwidth,clip=true,angle=270]{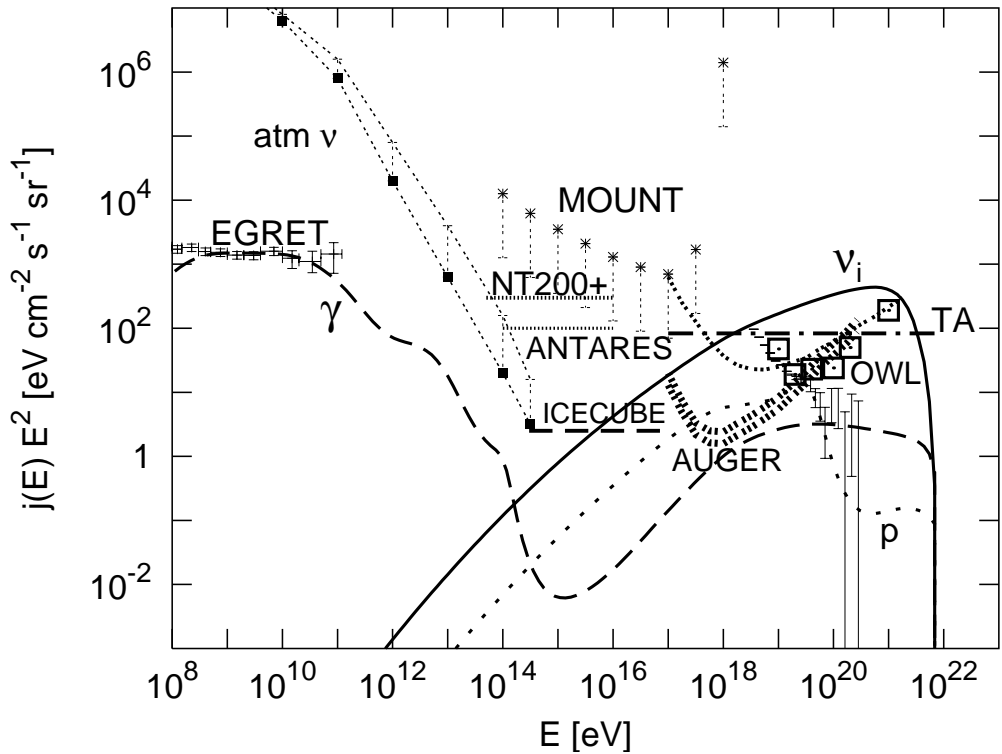}
\caption[...]{\cite{dima}~Predictions for the differential fluxes of
$\gamma-$rays (dashed line), nucleons (dotted line), and neutrinos
per flavor (solid line, assuming maximal mixing among all flavors)
in a top-down model characterized by $p=1$, $m_X = 2\times10^{14}\,$GeV,
and the decay mode $X\to q+q$, assuming the FF in MLLA
without supersymmetry~\cite{detal}, with a fraction of
10\% nucleons. The calculation used the code described in Ref.~\cite{code}
and assumed the minimal URB version consistent with observations~\cite{PB}
and an EGMF of $10^{-12}\,$G. Measured CR, neutrino, and $\gamma-$ray
data are as in Fig.~\ref{fig6} (however, only AGASA UHECR data are
shown). Also shown are
expected sensitivities of the currently being constructed
Auger project to electron/muon and tau-neutrinos~\cite{auger_nu},
and the planned projects telescope array (TA)~\cite{ta_nu}, the
fluorescence/\v{C}erenkov detector MOUNT~\cite{mount}, and the
space based OWL~\cite{owl} (we take the latter as representative
also for EUSO~\cite{euso}), the water-based NT200+~\cite{baikal_limit},
ANTARES~\cite{antares}, and the ice-based ICECUBE~\cite{icecube}, as
indicated.
\label{fig7}}
\end{figure}

Fig.~\ref{fig7} shows results for the time averaged nucleon, $\gamma-$ray,
and neutrino fluxes in a typical top-down scenario, along with
constraints on diffuse $\gamma-$ray fluxes at GeV energies
and neutrino flux sensitivities of
future experiments. The spectrum was optimally normalized 
to allow for an explanation of the observed UHECR
events, assuming their consistency with a nucleon or
$\gamma-$ray primary. The flux below $\la2\times10^{19}\,$eV
is presumably due to conventional
acceleration in astrophysical sources and was not fit.
The PP process on the CMB depletes the photon flux above 100 TeV, and the
same process on the IR/O background causes depletion of the photon flux in
the range 100 GeV--100 TeV, recycling the absorbed fluxes to
energies below 100 GeV through EM cascading.
The predicted background is {\it not} very sensitive to
the specific IR/O background model, however~\cite{ahacoppi}.
The scenario in Fig.~\ref{fig7} obviously
obeys all observational constraints within the normalization
ambiguities. Note
that the diffuse $\gamma-$ray background measured by
EGRET~\cite{egret} up to 10 GeV puts a strong constraint on these
scenarios, especially if there is already a significant
contribution to this background from conventional
sources such as unresolved $\gamma-$ray blazars~\cite{muk-chiang}.
However, this constraint is much weaker for TDs or decaying long
lived X particles with a non-uniform clustered density~\cite{bkv}.

The energy loss and absorption lengths for UHE nucleons and photons
are $\la100$ Mpc. Thus, predicted UHE nucleon and photon fluxes are
independent of cosmological evolution. The $\gamma-$ray flux
below $\simeq10^{11}\,$eV, however, scales as the
total X particle energy release integrated over all redshifts
and increases with decreasing $p$~\cite{sjsb}. For
$m_X=2\times10^{16}\,$GeV, scenarios with $p<1$ are therefore ruled
out, whereas models with a comovingly constant injection rate
($p=2$) are well within the limits.

It is clear from the above discussions that the predicted particle fluxes
in the top-down scenarios are currently uncertain to a large extent due to 
particle physics uncertainties (e.g., mass and decay modes of the X
particles, the quark FF, the nucleon fraction $f_N$,
and so on) as well as astrophysical uncertainties (e.g., strengths of the
radio and infrared backgrounds, extragalactic magnetic fields, etc.). 
More details on the dependence of the predicted UHE particle spectra and
composition on these particle physics and astrophysical
uncertainties are contained in Ref.~\cite{SLBY}.

We stress here that there are viable top-down scenarios which
predict nucleon fluxes comparable to or even higher than
the $\gamma-$ray flux at all energies, even though $\gamma-$rays
dominate at injection. This occurs, e.g., in the case of high URB
and/or for a strong EGMF, and a nucleon fraction of $\simeq10\%$. Some
of these top-down scenarios would therefore remain viable even if
UHECR induced EAS should be proven inconsistent with photon
primaries~\cite{gamma_comp}). This is in contrast to scenarios with
decaying massive dark matter in the Galactic halo which,
due to the lack of interactions of injected particles, predict
compositions directly given by the FFs, i.e. domination by
$\gamma-$rays, and thus may be in conflict with observed
compositions~\cite{gamma_comp}.

The normalization procedure to the UHECR flux described above
imposes the constraint $Q^0_{\rm UHECR}\la10^{-22}\,{\rm eV}\,{\rm
cm}^{-3}\,{\rm sec}^{-1}$ within a factor of a
few~\cite{ps1,SLBY,slsc} for the total energy release rate $Q_0$
from TDs at the current epoch.
In most top-down models, because of the unknown values of the
parameters involved, it is currently not
possible to calculate the exact value of $Q_0$ from first principles,
although it has been shown that the required values of $Q_0$ (in order to
explain the UHECR flux) mentioned above are quite possible for
certain kinds of TDs. Some cosmic
string simulations and the necklace scenario suggest that
defects may lose most of
their energy in the form of X particles and estimates of this
rate have been given~\cite{vincent,BV}. If that is the case, the
constraint on $Q^0_{\rm UHECR}$ translates via Eq.~(\ref{dotnx})
into a limit on the symmetry
breaking scale $\eta$ and hence on the mass $m_X$ of the X particle: 
$\eta\sim m_X\la10^{13}\,$GeV~\cite{wmgb}. Independently 
of whether or not this scenario explains UHECR, the EGRET measurement
of the diffuse GeV $\gamma-$ray background leads to a similar bound,
$Q^0_{\rm EM}\la2.2\times10^{-23}\,h
(3p-1)\,{\rm eV}\,{\rm cm}^{-3}\,{\rm sec}^{-1}$, which leaves
the bound on $\eta$ and $m_X$ practically unchanged.
Furthermore, constraints from limits on CMB distortions and light
element abundances from $^4$He-photo-disintegration are
comparable to the bound from the directly observed
diffuse GeV $\gamma$-rays~\cite{sjsb}. That these crude
normalizations lead to values of $\eta$ in the right range
suggests that defect models require less fine tuning than
decay rates in scenarios of metastable massive dark matter.
Furthermore, extragalactic top-down models would probably be
less motivated, but not be ruled out automatically if the
existence of the GZK cutoff is confirmed~\cite{hires}, in contrast
to decaying massive dark matter scenarios.

As discussed above, in top-down scenarios most of the energy is
released in the form of EM particles and neutrinos. If the X
particles decay into a quark and a lepton, the quark hadronizes
mostly into pions and the ratio of energy release into the
neutrino versus EM channel is $r\simeq0.3$. The energy fluence
in neutrinos and $\gamma-$rays is thus comparable. However,
whereas the photons are recycled down to the GeV range where
their flux is constrained by the EGRET measurement, the neutrino
flux is practically not changed during propagation and thus
reflects the injection spectrum. Its predicted level is consistent
with all existing upper limits (compare Fig.~\ref{fig7} with
Fig.~\ref{fig6}) but should be detectable by several experiments
under construction or in the proposal stage (see Fig.~\ref{fig7}).
This would allow to directly see the quark fragmentation spectrum.

\section{Experimental Projects and the Future}
The field of ultra-high energy cosmic radiation is certainly
experimentally driven. This is obvious from the lack of convergence
of theoretical models caused by the sparseness of yet available
data. To demonstrate the prospects for a considerable increase
of data we therefore conclude with a short discussion of experimental
projects.

An upscaled version of the old Fly's Eye experiment, the
High Resolution Fly's Eye (HiRes) detector is looking for CRs
in Utah, USA~\cite{hires}. Taking into account a duty cycle of about
10\% because a fluorescence detector requires clear, moonless nights,
this instrument will collect events above $10^{17}\,$eV at a rate
about 10 times larger than the old Fly's Eye, corresponding
to a few events above $10^{20}\,$eV per year.
The largest project presently under construction is the Pierre Auger
Giant Array Observatory~\cite{auger} planned for two sites, one in
Mendoza, Argentina and another in Utah, USA for maximal sky coverage.
Each site will have a $\simeq3000\,{\rm km}^2$ ground array. The southern
site will have about 1600 particle detectors (separated by 1.5 km
each) overlooked by four fluorescence detectors. The ground arrays
will have a duty cycle of nearly 100\%,
leading to detection rates about 30 times as large as for the AGASA
array, i.e. about 50 events per year above $10^{20}\,$eV. About 10\%
of the events will be detected simultaneously by the ground array
and the fluorescence component and can be used for cross
calibration and detailed EAS studies. The detection energy threshold will
be around $10^{18}\,$eV. These instruments will also have
considerable sensitivity to neutrinos above $\sim10^{19}\,$eV,
typically from the near-horizontal air-showers that are produced
by them~\cite{auger_nu}, as shown in Fig.~\ref{fig7}. The old Fly's Eye
experiment~\cite{baltrusaitis} and the AGASA experiment~\cite{agasa_nu}
have already established upper limits on neutrino fluxes based on the
non-observation of horizontal air showers, see Fig.~\ref{fig6}.

As Fig.~\ref{fig7} demonstrates for the top-down models, neutrino
fluxes in the intermediate energy range from $\simeq10^{12}\,$eV
to $\simeq10^{15}\,$eV can also are relevant for certain models
including active galactic nuclei as CR sources. Traditionally,
neutrinos in this range are detected via the EM showers or the
\v{C}erenkov light induced by the leptons produced by charged
current reactions in ice or water. The largest operative experiment
using ice is the Antarctic Muon And Neutrino Detector Array
(AMANDA)~\cite{amanda} at the South pole which also has
a next generation version named ICECUBE~\cite{icecube} which may
reach up to $\simeq10^{18}\,$eV. A water based version of this
technique is operative in Lake Baikal~\cite{baikal} and another
experiment for Astronomy with a Neutrino Telescope and Abyss
environmental RESarch (ANTARES) is under
construction in the Mediterranean~\cite{antares}.

In addition, there are plans to construct telescopes to detect
fluorescence and \v{C}erenkov light from near-horizontal showers produced in
mountain targets by neutrinos in the intermediate window of
energies between $\sim10^{15}\,$eV and $\sim10^{19}\,$eV~\cite{fargion,mount}.
The alternative of detecting neutrino and CR induced showers by
triggering onto the radio pulses emitted by them is also investigated
currently~\cite{radhep}. Two implementations of this technique,
the Radio Ice \v{C}erenkov Experiment (RICE), a small array of radio
antennas in the South pole ice~\cite{rice},
and the Goldstone Lunar Ultra-high energy neutrino Experiment (GLUE),
based on monitoring of the moons rim with the NASA Goldstone
radio telescope for radio pulses from neutrino-induced
showers~\cite{glue}, have so far produced neutrino flux upper limits.
It may be possible to use future large scale projects for
radio telescopes such as the LOw Frequency ARray (LOFAR) for UHECR
and neutrino studies of low energy and angular resolution but high
statistics~\cite{fg}.
Acoustic detection of neutrino induced interactions is also
being considered~\cite{acoustic}.

There are also plans to detect EAS in the Earth's atmosphere from
space. This would provide an increase by another factor $\sim50$ in
collecting power compared to the Pierre Auger Project, i.e. an
event rate above $10^{20}\,$eV of up to a few thousand per year.
Two concepts are currently being studied, the Orbiting Wide-angle
Light-collector (OWL)~\cite{owl} in the USA and the Extreme Universe
Space Observatory (EUSO)~\cite{euso} in Europe of which a prototype may
fly on the International Space Station.

Space-based detectors would be especially suitable for detection
of very small event rates such as those caused by neutrino
primaries which rarely interact in the atmosphere due to their
small interaction cross sections. This disadvantage for the detection
process is at the same time a blessing because it makes these elusive
particles reach us unattenuated over cosmological distances and from
very dense environments where all other particles (except gravitational
waves) would be absorbed. Giving rise to showers typically starting
deep within the atmosphere, they can also be distinguished
from other primaries.

We believe that these experimental developments will allow to test,
constrain, or rule out many of the theoretical scenarios and speculations
discussed here. This will begin already within the next few years.

\section*{Acknowledgments}
I would like to thank all my collaborators without whose contributions
and help this review would have been impossible.


\begin{thebibliography}{99}

\bibitem{hess} V.~F.~Hess, Phys.~Z. 13 (1912) 1084.

\bibitem{auger_disc} P.~Auger, R.~Maze, T.~Grivet-Meyer,
{\it Acad\'emie des Sciences} 206 (1938) 1721; P.~Auger, R.~Maze,
{\it ibid.} 207 (1938) 228.

\bibitem{crbook} for a general introduction on cosmic rays
see, e.g., V.~S.~Berezinsky, S.~V.~Bulanov, V.~A.~Dogiel,
V.~L.~Ginzburg, V.~S.~Ptuskin, {\it Astrophysics of Cosmic
Rays} (North-Holland, Amsterdam, 1990); T.~K.~Gaisser, {\it
Cosmic Rays and Particle Physics}, Cambridge University Press
(Cambridge, 1998).

\bibitem{battiston} for a recent overview see, e.g., R.~Battiston,
e-print astro-ph/0208108.

\bibitem{haverah} See, e.g., M.~A.~Lawrence, R.~J.~O.~Reid, and
A.~A.~Watson, J.~Phys.~G Nucl.~Part.~Phys. 17 (1991) 733, and
references therein; see also
{\sf http://ast.leeds.ac.uk/haverah/hav-home.html}.

\bibitem{agasa} M.~Takeda et al., Phys. Rev. Lett. 81 (1998) 1163;
Astrophys. J. 522 (1999) 225; Hayashida et al.,
e-print astro-ph/0008102; see also
{\sf http~://www-akeno.icrr.u-tokyo.ac.jp/AGASA/}.

\bibitem{fe} D.~J.~Bird et al., Phys.~Rev.~Lett. 71 (1993)
3401; Astrophys.~J. 424 (1994) 491; ibid. 441 (1995) 144.

\bibitem{hires} T.~Abu-Zayyad et al. (HiRes collaboration),
e-print astro-ph/0208243; e-print astro-ph/0208301.

\bibitem{reviews} for recent reviews see J.~W.~Cronin, Rev.~Mod.~Phys.
71 (1999) S165; M.~Nagano, A.~A.~Watson, Rev.~Mod.~Phys. 72 (2000) 689;
A.~V.~Olinto, Phys.~Rept. 333-334 (2000) 329; X.~Bertou,
M.~Boratav, and A.~Letessier-Selvon, Int.~J.~Mod.~Phys. A15 (2000) 2181;
G.~Sigl, Science 291 (2001) 73.

\bibitem{bs-rev} P.~Bhattacharjee and G.~Sigl,
Phys.~Rept. 327 (2000) 109; L.~Anchordoqui, T.~Paul, S.~Reucroft,
and J.~Swain, e-print hep-ph/0206072.

\bibitem{school} ``Physics and Astrophysics of Ultra High Energy Cosmic Rays'',
Lecture Notes in Physics, vol.~576 (Springer Verlag, 2001),
eds. M.~Lemoine, G.~Sigl.

\bibitem{hillas-araa} A.~M.~Hillas, Ann.~Rev.~Astron.~Astrophys. 22
(1984) 425. 

\bibitem{ssb} G.~Sigl, D.~N.~Schramm, and P.~Bhattacharjee,
Astropart.~Phys. 2 (1994) 401.

\bibitem{norman} C.~A.~Norman, D.~B.~Melrose, and A.~Achterberg,
Astrophys.~J. 454 (1995) 60. 

\bibitem{gzk} K.~Greisen, Phys.~Rev.~Lett. 16 (1966)
748; G.~T.~Zatsepin and V.~A.~Kuzmin, Pis'ma
Zh. Eksp. Teor. Fiz. 4 (1966) 114 [JETP. Lett. 4 (1966) 78].

\bibitem{bbo} see, e.g., M.~Blanton, P.~Blasi, and A.~V.~Olinto,
Astropart.~Phys. 15 (2001) 275.

\bibitem{heavy} J.~L.~Puget, F.~W.~Stecker, and J.~H.~Bredekamp,
Astrophys. J. 205 (1976) 638;
L.~N.~Epele and E.~Roulet, Phys.~Rev.~Lett. 81
(1998) 3295; J.~High Energy~Phys. 9810 (1998) 009;
F.~W.~Stecker, Phys.~Rev.~Lett. 81 (1998) 3296;
F.~W.~Stecker and M.~H.~Salamon, Astrophys.~J.
512 (1999) 521.

\bibitem{elb-som} J.~W.~Elbert, and P.~Sommers, Astrophys.~J. 441
(1995) 151.

\bibitem{tt} P.~G.~Tinyakov and I.~I.~Tkachev,
Pisma~Zh.~Eksp.~Teor.~Fiz. 74 (2001) 3 [JETP Lett. 74 (2001) 1].

\bibitem{wb} for a discussion see, e.g., J.~N.~Bahcall and E.~Waxman,
e-print hep-ph/0206217.

\bibitem{auger} J.~W.~Cronin, Nucl.~Phys.~B (Proc.~Suppl.) 28B (1992)
213; The Pierre Auger Observatory Design Report (ed.~2), March 1997;
see also {\sf http://www.auger.org}.

\bibitem{nu_review} for a recent review, see F.~Halzen and D.~Hooper,
Rept.~Prog.~Phys. 65 (2002) 1025.

\bibitem{gammarev} R.~A.~Ong, Phys.~Rept. 305 (1998) 95;
M.~Catanese and T.~C.~Weekes, e-print astro-ph/9906501,
invited review, Publ.~Astron.~Soc. of the Pacific, Vol.~111,
issue 764 (1999) 1193.

\bibitem{wb-bound} E.~Waxman and J.~Bahcall, Phys.~Rev.~D. 59 (1999)
023002; J.~Bahcall and E.~Waxman, Phys.~Rev.~D 64 (2001) 023002.

\bibitem{texas19} Proc. {\it 19$^{th}$ Texas Symposium on
Relativistic Astrophysics}, Paris (France), eds. E. Aubourg, et al.,
Nuc.~Phys.~B~(Proc.~Supp.) 80B (2000).

\bibitem{mpr} K.~Mannheim, R.~J.~Protheroe, J.~P.~Rachen,
Phys.~Rev.~D 63 (2001) 023003; J.~P.~Rachen, R.~J.~Protheroe,
K.~Mannheim, astro-ph/9908031, in Ref.~\cite{texas19}.

\bibitem{sigl} for more details see, e.g., G.~Sigl, in Ref.~\cite{school},
pp.~196-254.

\bibitem{code} S.~Lee, Phys. Rev. D 58 (1998) 043004;
O.~E.~Kalashev, V.~A.~Kuzmin, and D.~V.~Semikoz,
e-print astro-ph/9911035; Mod.~Phys.~Lett A16 (2001) 2505.

\bibitem{kkss1} O.~E.~Kalashev, V.~A.~Kuzmin, D.~V.~Semikoz, and
G.~Sigl, Phys.~Rev.~D 65 (2002) 103003.

\bibitem{kkss2} O.~E.~Kalashev, V.~A.~Kuzmin, D.~V.~Semikoz, and
G.~Sigl, Phys.~Rev.~D 66 (2002) 063004.

\bibitem{zburst1} T.~J.~Weiler, Phys.~Rev.~Lett. 49 (1982) 234;
Astrophys.~J. 285 (1984) 495;
E.~Roulet, Phys.~Rev.~D 47 (1993) 5247;
S.~Yoshida, Astropart.~Phys. 2 (1994) 187.

\bibitem{zburst2} T.~J.~Weiler, Astropart.~Phys. 11 (1999) 317;
D.~Fargion, B.~Mele, and A.~Salis, Astrophys.~J.
517 (1999) 725; S.~Yoshida, G.~Sigl, and S.~Lee, Phys.~Rev.~Lett. 81
(1998) 5505; Z.~Fodor, S.~D.~Katz, and A.~Ringwald,
Phys.~Rev.~Lett. 88 (2002) 171101.

\bibitem{egret} P.~Sreekumar et al., Astrophys.~J. 494 (1998) 523.

\bibitem{bko} V.~Berezinsky, M.~Kachelrie\ss\ ,and S.~Ostapchenko,
e-print hep-ph/0205218.

\bibitem{Clark} T.~A.~Clark, L.~W.~Brown, and J.~K.~Alexander, Nature
228 (1970) 847.

\bibitem{PB} R.~J.~Protheroe and P.~L.~Biermann,
Astropart. Phys. 6 (1996) 45.

\bibitem{irb}
J.~R.~Primack, R.~S.~Somerville, J.~S.~Bullock, and J.~E.~Devriendt,
%``Probing galaxy formation with high energy gamma-rays,''
e-print astro-ph/0011475, AIP~Conf.~Proc. 558 (2001) 463.
%%CITATION = ASTRO-PH 0011475;%%

\bibitem{bt_review} for a review see, e.g., D.~Grasso and
H.~Rubinstein, Phys.~Rept. 348 (2001) 163.

\bibitem{bo_review} J.~P.~Vall{\'e}e, Fundamentals of Cosmic
Physics, Vol.~19 (1997) 1; J.-L.~Han and R.~Wielebinski,
e-print astro-ph/0209090.

\bibitem{ryu}D.~Ryu, H.~Kang, and P.~L.~Biermann,
Astron.~Astrophys. 335 (1998) 19.

\bibitem{blasi} P.~Blasi, S.~Burles, and A.~V.~Olinto, Astrophys.~J.
514 (1999) L79.

\bibitem{mte} G.~Medina-Tanco and T.~En\ss lin, Astropart.~Phys.
16 (2001) 47.

\bibitem{wm} E.~Waxman and J.~Miralda-Escud\'{e}:
Astrophys.~J.472 (1996) L89.

\bibitem{is} C.~Isola and G.~Sigl, e-print astro-ph/0203273.

\bibitem{miniati} F.~Miniati, e-print astro-ph/0203014.

\bibitem{ems} T.~En\ss lin, F.~Miniati, and G.~Sigl, in preparation.

\bibitem{ames} J.~Alvarez-Mu\~niz, R.~Engel, and T.~Stanev,
Astrophys.~J. 572 (2001) 185.

\bibitem{bils} G.~Bertone, C.~Isola, M.~Lemoine, and
G.~Sigl, e-print astro-ph/0209192.

\bibitem{ggs} D.~S.~Gorbunov, G.~G.~Raffelt, and D.~V.~Semikoz,
Phys.~Rev.~D 64 (2001) 096005.

\bibitem{cfk} G.~R.~Farrar, Phys.~Rev.~Lett. 76 (1996) 4111;
D.~J.~H.~Chung, G.~R.~Farrar, and E.~W.~Kolb, Phys.~Rev.~D
57 (1998) 4696.

\bibitem{gluino} I.~F.~Albuquerque et al. (E761 collaboration),
Phys.~Rev.~Lett. 78 (1997) 3252; A.~Alavi-Harati et al.
(KTeV collaboration), Phys.~Rev.~Lett. 83 (1999) 2128.

\bibitem{bllac} P.~G.~Tinyakov and I.~I.~Tkachev, JETP Lett.
74 (2001) 445; D.~S.~Gorbunov, P.~G.~Tinyakov, I.~I.~Tkachev,
and S.~V.~Troitsky, e-print astro-ph/0204360.

\bibitem{gqrs} R.~Gandhi, C.~Quigg, M.~H.~Reno, and I.~Sarcevic,
Astropart.~Phys. 5 (1996) 81; Phys.~Rev.~D 58 (1998) 093009.

\bibitem{kovesi-domokos} G.~Domokos and S.~Kovesi-Domokos,
Phys.~Rev.~Lett. 82 (1999) 1366.

\bibitem{tev-qg} N.~Arkani-Hamed, S.~Dimopoulos, and
G.~Dvali, Phys.~Lett. B 429 (1998) 263; I.~Antoniadis,
N.~Arkani-Hamed, S.~Dimopoulos, and G.~Dvali, Phys.~Lett. B
436 (1998) 257; N.~Arkani-Hamed, S.~Dimopoulos, and G.~Dvali,
Phys.~Rev.~D 59 (1999) 086004.

\bibitem{aco} E.-J.~Ahn, M.~Cavaglia, and A.~V. Olinto,
e-print hep-th/0201042.

\bibitem{fs} J.~L.~Feng and A.~D.~Shapere, Phys.~Rev.~Lett.
88 (2002) 021303.

\bibitem{kp} M.~Kachelrie\ss\ and M.~Pl\"umacher, Phys.~Rev.~D
62 (2000) 103006.

\bibitem{dima} I acknowledge Dmitry Semikoz for helping me with
these figures which are based on work in Ref.~\cite{kkss2}.

\bibitem{lipari} see, e.g., 
P.~Lipari,
%``Lepton Spectra In The Earth's Atmosphere,''
Astropart.\ Phys.\  {\bf 1}, 195 (1993).
%%CITATION = APHYE,1,195;%%

\bibitem{macro} For general information see
{\sf http://wsgs02.lngs.infn.it:8000/macro/}; see
also M.~Ambrosio {\it et al.}  [MACRO Collaboration],
%``Search for diffuse neutrino flux from astrophysical sources with MACRO,''
astro-ph/0203181.
%%CITATION = ASTRO-PH 0203181;%%

\bibitem{amanda_limit} J.~Ahrens et al., AMANDA collaboration,
e-print hep-ph/0112083, Proceedings of the EPS International
Conference on High Energy Physics, Budapest, 2001 (D. Horvath, P. Levai, A.
Patkos, eds.), JHEP Proceedings Section, PrHEP-hep2001/207.
%%CITATION = HEP-PH 0112083;%%

\bibitem{baikal_limit} V.~Balkanov {\it et al.}  [BAIKAL Collaboration],
%``Baikal experiment: Status report,''
astro-ph/0112446.
%%CITATION = ASTRO-PH 0112446;%%

\bibitem{agasa_nu} S. Yoshida for the AGASA Collaboration,
Proc. of 27th ICRC (Hamburg) 3 (2001) 1142.

\bibitem{baltrusaitis} R.~M.~Baltrusaitis et al.,
Astrophys.~J. 281 (1984) L9; Phys.~Rev.~D 31 (1985) 2192.

\bibitem{rice} I.~Kravchenko et al. (RICE collaboration), e-print
astro-ph/0206371; for general information on
RICE see {\sf http://kuhep4.phsx.ukans.edu/~iceman/index.html}.

\bibitem{glue} P.~W.~Gorham, K.~M.~Liewer, C.~J.~Naudet,
e-print astro-ph/9906504, Proc. of the {\it 26th International Cosmic
Ray Conference (ICRC 99), Salt Lake City, Utah}, Vol. 2, p.~479;
P.~W.~Gorham et al., e-print astro-ph/0102435.

\bibitem{mr} D.~A.~Morris and A.~Ringwald, Astropart.~Phys. 2 (1994)
43.

\bibitem{tol} C.~Tyler, A.~Olinto, and G.~Sigl, Phys.~Rev.~D 63 (2001) 055001.

\bibitem{afgs} see, e.g., L.~A.~Anchordoqui, J.~L.~Feng, H.~Goldberg, and
A.~D.~Shapere, Phys.~Rev.~D 65 (2002) 124027; e-print hep-ph/0207139.

\bibitem{kw} A.~Kusenko and T.~Weiler, Phys.~Rev.~Lett. 88 (2002)
161101.

\bibitem{dgln} G.~Dvali, G.~Gabadadze, M.~Kolanovi\'{c}, and
F.~Nitti, Phys.~Rev.~D 65 (2001) 024031.

\bibitem{dgs} G.~Dvali, G.~Gabadadze, and M.~Shifman,
e-print hep-th/0202174.

\bibitem{sigl1} G.~Sigl, e-print hep-ph/0207254.

\bibitem{vli_others} H.~Sato and T.~Tati, Prog.~Theor.~Phys. 47 (1972)
1788; D.~A.~Kirzhnits and V.~A.~Chechin, Sov.~J.~Nucl.~Phys.
15 (1972) 585; L.~Gonzalez-Mestres, e-print hep-th/0208064.

\bibitem{cg} S.~Coleman and S.~L.~Glashow, Phys.~Lett. B405 (1997)
249; Phys.~Rev.~D 59 (1999) 116008.

\bibitem{amelino-piran} G.~Amelino-Camelia and T.~Piran:,
Phys.~Lett. B497 (2001) 265.

\bibitem{ms} J.~Magueijo and L.~Smolin, e-print gr-qc/0207085.

\bibitem{ck} D.~Colladay and V.~Kostelecky, Phys.~Rev.~D 59 (1999) 116002.

\bibitem{emn} see, e.g., J.~Ellis, N.~E.~Mavromatos, and D.~V.~Nanopoulos,
Phys.~Rev.~D 62 (2000) 084019.

\bibitem{aloisio} R.~Aloisio, P.~Blasi, P.~Ghia, and A.~Grillo,
Phys.~Rev.~D 62 (2000) 053010.

\bibitem{tev} see, e.g., R.~Protheroe and H.~Meyer, Phys.~Lett.
B493 (1996) 1; S.~Liberati, T.~Jacobson, and D.~Mattingly, e-print
hep-ph/0112207.

\bibitem{jlm} T.~Jacobson, S.~Liberati, and D.~Mattingly, e-print
hep-ph/0209264.

\bibitem{kuz-tak} for a brief review see V.~Kuzmin and I.~Tkachev,
Phys.~Rept. 320 (1999) 199

\bibitem{kb} V.~Berezinsky, M.~Kachelrie\ss\ , and A.~Vilenkin,
Phys.~Rev.~Lett. 79 (1997) 4302.

\bibitem{BHS} P.~Bhattacharjee, C.~T.~Hill, and D.~N.~Schramm,
Phys.~Rev.~Lett. 69 (1992) 567.

\bibitem{BR} see, e.g., P.~Bhattacharjee and N.~C.~Rana,
Phys.~Lett.~B 246 (1990) 365.

\bibitem{BV} V.~Berezinsky and A.~Vilenkin, Phys.~Rev.~Lett.
79 (1997) 5202.

\bibitem{BS} P.~Bhattacharjee and G.~Sigl, Phys.~Rev.~D 51 (1995)
4079 .

\bibitem{detal} Yu.~L.~Dokshitzer, V.~A.~Khoze, A.~H.~M\"uller,
and S.~I.~Troyan, {\sl Basics of
Perturbative QCD} (Editions Frontieres, Singapore, 1991).

\bibitem{frag} see, e.~g., S.~Sarkar and R.~Toldr\`{a},
Nucl.~Phys.~B 621 (2002) 495; C.~Barbot and M.~Drees,
Phys.~Lett.~B 533 (2002) 107.

\bibitem{auger_nu}
J.~J.~Blanco-Pillado, R.~A.~V\'{a}zquez, and
E.~Zas, Phys.~Rev.~Lett. 78 (1997) 3614;
K.~S.~Capelle, J.~W.~Cronin, G.~Parente, and
E.~Zas, Astropart.~Phys. 8 (1998) 321;  A.~Letessier-Selvon,
e-print astro-ph/0009444, AIP~Conf.~Proc 566 (2000) 157; X.~Bertou
et al., Astropart.~Phys. 17 (2002) 183.

\bibitem{ta_nu} M.~Sasaki and M.~Jobashi, e-print astro-ph/0204167.

\bibitem{mount} G.~W.~S.~Hou and M.~A.~Huang, astro-ph/0204145.

\bibitem{owl} D.~B.~Cline and F.~W.~Stecker, OWL/AirWatch science
white paper, e-print astro-ph/0003459.

\bibitem{euso} See {\sf http://www.ifcai.pa.cnr.it/Ifcai/euso.html}.

\bibitem{antares} For general information see
{\sf http://antares.in2p3.fr}; see
also S.~Basa, (e-print astro-ph/9904213, in~\cite{texas19}; 
ANTARES Collaboration, e-print astro-ph/9907432. 

%\bibitem{amanda-II} R.~Wischnewski (AMANDA Collaboration),
%e-print astro-ph/0204268.

\bibitem{icecube} For general information see
{\sf http://www.ps.uci.edu/$\sim$icecube/workshop.html}; see also
F.~Halzen: Am.~Astron.~Soc. Meeting 192, \# 62 28
(1998); AMANDA collaboration: astro-ph/9906205, Proc. {\it 8$^{th}$
International Workshop on Neutrino Telescopes}, Venice, Feb. 1999.

\bibitem{ahacoppi} P.~S.~Coppi and F.~A.~Aharonian,
Astrophys.~J. 487 (1997) L9.

\bibitem{muk-chiang} R.~Mukherjee and J.~Chiang,
Astropart.~Phys. 11 (1999) 213.

\bibitem{bkv} V.~Berezinsky, M.~Kachelrie\ss\ , and A.~Vilenkin,
Phys.~Rev.~Lett. 79 (1997) 4302.

\bibitem{sjsb} G.~Sigl, K.~Jedamzik, D.~N.~Schramm, and
V.~Berezinsky, Phys.~Rev.~D 52 (1995) 6682.

\bibitem{SLBY} G.~Sigl, S.~Lee, P.~Bhattacharjee, and S.~Yoshida, 
Phys.~Rev.~D 59 (1999) 043504.

\bibitem{gamma_comp} M.~Ave et al., Phys.~Rev.~D 65 (2002) 063007;
K.~Shinozaki et al. (AGASA collaboration), Astrophys.~J. 571 (2002)
L117.

\bibitem{ps1} R.~J.~Protheroe and T.~Stanev, Phys.~Rev.~Lett. 77
(1996) 3708; erratum, ibid. 78 (1997) 3420.

\bibitem{slsc} G.~Sigl, S.~Lee, D.~N.~Schramm, and P.~S.~Coppi,
Phys.~Lett.~B 392 (1997) 129.

\bibitem{vincent} G.~R.~Vincent, N.~D.~Antunes, and
M.~Hindmarsh, Phys.~Rev.~Lett. 80 (1998) 2277; G.~R.~Vincent,
M.~Hindmarsh, and M.~Sakellariadou, Phys.~Rev.~D 56 (1997) 637.

\bibitem{wmgb} U.~F.~Wichoski, J.~H.~MacGibbon, and R.~H.~Brandenberger,
Phys.~Rev.~D 65 (2002) 063005.

\bibitem{amanda} For general information see
{\sf http://amanda.berkeley.edu/};
see also F.~Halzen: New~Astron.~Rev 42 (1999) 289;
%%CITATION = NONE;%%
for recent physics results see, e.g., J.~Ahrens et al.
(AMANDA Collaboration), e-print astro-ph/0206487; e-print 
astro-ph/0208006.

\bibitem{baikal} For general information see
{\sf http://www-zeuthen.desy.de/baikal/baikalhome.html}; see also
V.~Balkanov et al (Baikal Collaboration), Proceedings of the IX
International Workshop on Neutrino Telescopes,
Venezia, March 6-9, 2001, Vol.II, p.591 (Ed. M.Balda-Ceolin). 

\bibitem{fargion} see, e.g.,
D.~Fargion,
%``Upward and horizontal tau airshowers by UHE nu,''
e-print hep-ph/0111289.
%%CITATION = HEP-PH 0111289;%%

\bibitem{radhep} Proceedings of {\it First International Workshop on
Radio Detection of High-Energy Particles}, Amer.~Inst.~of~Phys., vol.~579
(2001),
%%CITATION = NONE;%%
and at {\sf http://www.physics.ucla.edu/~moonemp/radhep/workshop.html};
see also K.~Green, J.~L.~Rosner, D.~A.~Suprun,
and J.~F.~Wilkerson, e-print astro-ph/0205046.

\bibitem{fg} H.~Falcke and P.~Gorham, e-print astro-ph/0207226.

\bibitem{acoustic} see, e.g.,
L.~G.~Dedenko, I.~M.~Zheleznykh, S.~K.~Karaevsky, A.~A.~Mironovich,
V.~D.~Svet and A.~V.~Furduev,
%``Prospects For Deep-Sea Acoustic Detection Of Neutrinos,''
Bull.\ Russ.\ Acad.\ Sci.\ Phys.\ 61 (1997) 469
[Izv.\ Ross.\ Akad.\ Nauk.\ 61 (1997) 593].

\end{thebibliography}
\end{document}